\documentclass[prb,notitlepage,showpacs]{revtex4}
\usepackage{graphicx}
\usepackage{amsmath}
\usepackage{amssymb}
\usepackage{color}
\usepackage{bm}

\definecolor{purple}{rgb}{0.8,0,0.6}

\newcommand{\TITLE}{Supercritical instability in graphene with two charged impurities}

\begin{document}

\title{\TITLE}

\date{\today}

\author{E.V. Gorbar}
\affiliation{Department of Physics, Taras Shevchenko National Kiev University, Kiev, 03022, Ukraine}
\affiliation{Bogolyubov Institute for Theoretical Physics, Kiev, 03680, Ukraine}

\author{V.P. Gusynin}
\affiliation{Bogolyubov Institute for Theoretical Physics, Kiev, 03680, Ukraine}

\author{O.O. Sobol}
\affiliation{Department of Physics, Taras Shevchenko National Kiev University, Kiev, 03022, Ukraine}

\begin{abstract}
We study the supercritical instability in gapped graphene with two charged impurities separated
by distance $R$ using the two-dimensional Dirac equation for electron quasiparticles.
Attention is paid to a situation when charges of impurities  are subcritical, whereas their
total charge exceeds a critical one. The critical distance $R_{\rm cr}$ in the system of two charged
centers is defined as that at which the electron  bound state with the lowest energy  reaches the
boundary of the lower continuum. A variational calculation of the critical distance $R_{\rm cr}$
separating the supercritical $(R<R_{\rm cr})$ and subcritical $(R>R_{\rm cr})$ regimes is carried out.
It is shown that the critical distance $R_{\rm cr}$ increases as the quasiparticle gap decreases.
The energy and width of a quasistationary state as functions of the distance between two
impurities are derived in the quasiclassical approximation.
\end{abstract}
\maketitle

\section{Introduction}

One of the most intriguing aspects of graphene physics is its deep and fruitful
relation with quantum electrodynamics (QED) and other quantum field theories.
The dynamics of the vacuum in QED leads to several peculiar effects not yet
observed in nature such as zitterbewegung (trembling motion), Klein tunneling, Schwinger pair
production, supercritical atomic collapse, and a new symmetry broken phase at strong coupling.
Theoretically, it was shown a long time ago\cite{Semenoff} that quasiparticle excitations
in graphene have a linear dispersion at low energies and are described by the massless
Dirac equation in $2+1$ dimensions. In the continuum limit, graphene model on a honeycomb
lattice maps onto a $2+1$-dimensional field theory of Dirac fermions interacting through
the $1/r$ Coulomb potential. Therefore, graphene could be used as a bench-top particle-physics
laboratory allowing us to investigate the fundamental interactions of matter. The Klein
tunneling  was observed experimentally in Ref.~[\onlinecite{Kim}] and quite recently
supercritical atomic collapse was observed for charged impurities in graphene in
Ref.~[\onlinecite{Wang}].

The instability of a supercritically charged impurity in graphene can be considered
as a condensed matter analog of atomic collapse in a strong Coulomb field.\cite{Greiner,Zeldovich} Theoretical works on the Dirac-Kepler problem in QED taking into account the finite size of nucleus\cite{finite-size} showed that for atoms with nuclear charge in excess of $Z> 170$
the electron states dive into the lower continuum leading to positron emission.\cite{Greiner,Zeldovich}
Since nuclei with such a large charge are not encountered in nature, the atomic collapse was
never observed in QED. Still supercritical fields can be temporarily created in a head-on
or nearly head-on collision of two very heavy nuclei.  The idea that the supercritical instability in QED can be experimentally tested in a collision of  heavy nuclei was suggested in the 1970s.\cite{Gershtein,Rafelski,Zeldovich} Subsequent experiments confirmed the existence
of supercritical fields in collisions of very heavy nuclei and the gross features of positron
emission,\cite{Greiner} however, an analysis of the supercritical regime turned out to be a difficult problem mainly due to the transient nature of supercritical fields generated during collisions.

From the viewpoint of the supercritical charge problem, the most remarkable feature of the
electron dynamics in graphene is that their effective Coulomb coupling with impurity with
charge $Ze$ is given by $\beta=Z\alpha/\kappa$, where $\alpha=e^2/\hbar v_{F}\simeq 2.19$ is the ``fine-structure'' coupling constant in graphene, $v_F \approx 10^6 {\rm m/s}$ is the velocity of
Dirac quasiparticles, and $\kappa$ is a dielectric constant. The very large value of coupling
constant compared to that in QED makes graphene an ideal platform for studying the supercritical
regime. The supercritical charge problem in gapless graphene was studied theoretically in
detail in Refs.~[\onlinecite{Shytov,Pereira}]. In the presence of a quasiparticle gap
$\Delta$, it was found that the critical coupling in a regularized Coulomb potential,
$V(r)=-\frac{Ze^{2}}{\kappa r}\theta(r-R_0)-\frac{Ze^{2}}{\kappa R_0}\theta(R_0-r)$,
is determined\cite{excitonic-instability} by $\beta_{c}=1/2+\pi^2/\log^2(c\Delta R_0/\hbar v_{F})$,
where $c\approx 0.21$.

The instability in the supercritical Coulomb center problem is closely related to the
excitonic instability in graphene in the supercritical coupling-constant regime
$\alpha>\alpha_c\sim 1$ (see Ref.~[\onlinecite{excitonic-instability}] and related papers\cite{Fertig,Guinea,VANT}).
In fact, the latter can be viewed as a many-body analog of the fall into the center phenomenon
and the critical coupling $\alpha_c$ is an analog of the critical coupling constant
$Z_ce^2/\hbar v_F$ in the problem of the Coulomb center.
The  quantum phase transition to the stable phase with excitonic (chiral) condensate and gapped quasiparticles may turn graphene into an insulator.\cite{metal-insulator,GGG2010,Gonzalez} This semimetal-insulator transition in graphene is widely discussed now in the literature,\cite{MS-phase-transition} it is similar to the chiral symmetry-breaking phase
transition that occurs in strongly coupled QED studied in the 1970s and 1980s (for a review,
see Ref.~[\onlinecite{reviews}]).
The predicted strong-coupling phase of QED was also searched in experiments in heavy-ion
collisions,\cite{Peccei} thus, like other QED effects not yet observed in nature, it has now
a chance to be tested in graphene.

Although, according to the theory, the supercritical instability should be easily realized
for charged impurities with $Z > 1$, its experimental observation remained elusive due to
the difficulty of producing highly charged impurities. However, one can reach the supercritical
regime by collecting a large enough number of charged impurities in a certain region. Recently,
this approach was successfully realized by creating artificial nuclei (clusters of charged
calcium dimers) on graphene\cite{Wang} using the tip of a scanning tunneling microscope.
It is ironic that in spite of a much larger value of coupling constant in graphene than in QED
the first observation of the supercritical instability in graphene still required the creation
of supercritical potentials from subcritical charges like in the case of heavy
nuclei collisions in QED discussed above. What crucially differs the graphene experiments\cite{Wang}
compared to that in QED is that the supercritical electric fields created by placing together
ionized Ca impurities are static unlike the fields created in heavy nuclei collisions in QED.
This makes it possible to observe and analyze reliably the supercritical regime.

In recent experiments\cite{Wang} atomic collapse was observed for a cluster of charged impurities
while existing theoretical studies considered only  a single multivalent Coulomb impurity.
To stay closer to the experimental situation, in the present paper we make a further step by
considering two Coulomb centers next to each other.  The main attention is paid to a situation
when the charges of impurities are subcritical, whereas their total charge exceeds a critical one.
We also include into consideration a quasiparticle gap that on the one side makes more transparent the derivation of the instability condition (diving of the lowest energy level into
the negative continuum), while on the other hand takes into account a possible presence
of a gap due to the interaction with a substrate.\cite{Zhou}

The paper is organized as follows. In Sec.~\ref{section-equation}, we set up the model,
introduce the notation, derive the asymptotics of the bound state solution that dives into
the lower continuum, and obtain an estimate of the critical distance between two charged
impurities for the onset of the supercritical regime. The energy and width of a quasistationary
state as  functions of the distance between two impurities are derived in Sec.~\ref{QS-state}
in  the quasiclassical approximation. A variational method is used in Sec.~\ref{variation-appr}
to find an improved expression for the critical distance as a function of the total charge
of impurities. The discussion of the results and conclusions are given in Sec.~\ref{section-conclusion}.
The Appendix at the end of the paper contains technical details and derivations used to supplement
the presentation in the main text.

\section{Dirac equation}
\label{section-equation}

The electron quasiparticle states in the vicinity of the $K_{\pm}$ points of graphene
in the field of two Coulomb impurities are described by the following Dirac Hamiltonian in $2+1$
dimensions (we set $\hbar=1$):
\begin{equation}
H=v_{F}\boldsymbol\tau\boldsymbol{p}+\xi\Delta\tau_z+V(r),
\label{Hamiltonian}
\end{equation}
where $\boldsymbol{p}$ is the canonical momentum, $\tau_{i}$ are the Pauli matrices, and $\Delta$
is a quasiparticle gap. The quasiparticle gap $\Delta$ can be generated if a graphene sheet is
placed on top of a substrate and two carbon sublattices become inequivalent because of interaction
with the substrate  (for band structure calculation of such a configuration see, for instance,
Ref.~[\onlinecite{Giovannetti}]). The gap can arise also in graphene ribbons due to geometrical
quantization\cite{Son} or due to many-body electron correlations.\cite{metal-insulator,GGG2010,Gonzalez,MS-phase-transition}

The Hamiltonian (\ref{Hamiltonian}) acts on two component spinor $\Psi_{\xi s}$ which carries the valley ($\xi=\pm)$ and spin ($s=\pm$) indices. We will use the standard convention:
$\Psi^{T}_{+s}=(\psi_{A},\psi_{B})_{K_{+}s}$, whereas $\Psi^{T}_{-s}=(\psi_{B},\psi_{A})_{K_{-}s}$, and $A,B$ refer to two sublattices of the hexagonal graphene lattice. The interaction potential of the electron
with two Coulomb impurities for $r_i > R_0$ ($i=1,2$) is given by
\begin{equation}
V\left(\mathbf{r}\right)=-\frac{e^2}{\kappa}\left(\frac{Z_1}{r_1}+\frac{Z_2}{r_2}\right),
\end{equation}
where $r_{1,2}=|\mathbf{r}\pm\mathbf{R}/2|$ measure distances from Coulomb impurities to the electron, $\kappa$ is the dielectric constant, and we assume that the Coulomb potential of each impurity is regularized by $-e^2 Z_i/(\kappa R_0)$ for $r_i < R_0$, where $R_0$ is of the order of graphene lattice
spacing. Since the interaction potential does not
depend on spin we will omit the spin index $s$ in what follows. Furthermore, for the sake of definiteness,
we will consider electrons in the $K_+$ valley (the Dirac equation for electrons in the $K_-$ valley
is obtained replacing by $\Delta$ with $-\Delta$). Since the experiments in Ref.~[\onlinecite{Wang}] were
performed for impurities of the same type, we will study in what follows the symmetric problem, i.e., $Z_1=Z_2=Z$. The main difficulty in solving the Dirac equation with two Coulomb centers in QED is that variables in this problem are not separable in any known orthogonal coordinate system.\cite{Popov} Unfortunately, this is true also for the Dirac equation for two Coulomb centers in the $(2+1)$-dimensional problem in graphene.

\subsection{Monopole approximation}

The Dirac equation for the electron in the potential of two charged impurities in graphene
\begin{equation}
\left(v_F\tau_xp_x+v_F\tau_yp_y+\Delta\tau_z+V\left(\mathbf{r}\right)\right)\Psi(\mathbf{r})
=E\Psi(\mathbf{r})
\label{Dirac-equation}
\end{equation}
for two-component spinor $\Psi(\mathbf{r})=(\phi,\ \chi)^T$ expressing $\chi$ in terms of $\phi$ gives the following second order
equation for the $\phi$ component of the Dirac spinor:
\begin{equation}
(\partial^2_x+\partial^2_y)\phi+\frac{\frac{\partial V}{\partial x}-i\frac{\partial V}{\partial y}}{E-V+\Delta}
\left(\frac{\partial\phi}{\partial x}+i\frac{\partial\phi}{\partial y}\right)+v^{-2}_F\left((E-V)^2-\Delta^2\right)\phi=0.
\label{upper-component}
\end{equation}

According to Refs.~[\onlinecite{Zeldovich,Greiner}], the supercritical instability takes place when
the bound state with the lowest energy dives into the lower continuum. This occurs when $E=-\Delta$.
For this solution, let us consider the asymptotic at large $r>>R$, where the potential equals
\begin{equation}
V\left(\mathbf{r}\right)=-\zeta v_F\left(\frac{1}{r}+\frac{R^2}{4r^3} P_2(\cos\varphi)+O\left(\frac{1}{r^5}\right)\right),
\label{potential-asymptotic}
\end{equation}
$\zeta=2Z\alpha/\kappa$ is a dimensionless charge, and $P_2(x)$ is the Legendre polynomial
$P_n(x)$ with $n=2$. In what follows we consider the case when charges of impurities  are subcritical whereas their total charge exceeds a critical one, $1/2<\zeta<1$. The case $\zeta<1/2$ corresponds
to the situation when the total charge is less than a critical one and is not considered in this paper.

Neglecting the quadrupole and higher order multipole terms in the potential (the monopole approximation) Eq. (\ref{upper-component}) reduces to the following equation for $\phi(r)$:
\begin{equation}
\label{e2}
\phi''+\frac{2}{r}\phi'+\left(\frac{\zeta^2}{r^2}-\frac{2m\zeta}{r}\right)\phi=0,
\end{equation}
where $m=\Delta/v^2_F$. The decreasing at infinity solution is expressed in terms of
a Macdonald function,
\begin{equation}
\phi(r)=C_1r^{-1/2}K_{i\gamma}(\sqrt{8m\zeta r}),\quad \gamma=\sqrt{4\zeta^2-1},
\label{phi-r-to-infinity}
\end{equation}
 with the asymptotic
\begin{equation}
\phi_{\rm asym}(r)=C_1r^{-3/4}\exp(-\sqrt{8m\zeta r}),\quad r \to \infty.
\label{asymptotic}
\end{equation}
This asymptotic is, of course, in agreement with the asymptotical behavior of a solution
for the Dirac equation of one center with charge $2Ze$. This shows also that the level that
reached the boundary of the lower continuum remains localized.

\subsection{An estimate of the critical distance}

\begin{figure}[ht]
  \centering
  \includegraphics[scale=0.5]{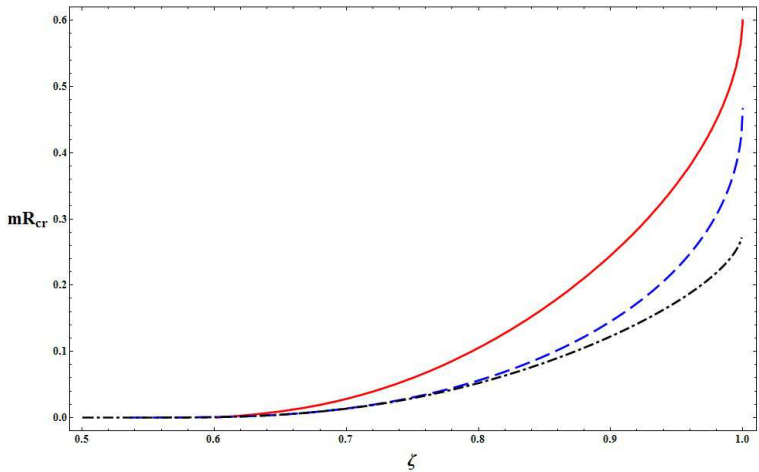}
  \caption{(Color online) The dependence $mR_{\rm cr}(\zeta)$ given by Eqs.~(\ref{trancend-eq})
  (dashed blue line) and (\ref{appr-num-solution}) (dash-dotted  black line), and solution of
  Eq.~(\ref{e4}) (solid red line). \label{appr-crit-line}}
\end{figure}

In order to find the asymptotic of the solution in the vicinity of Coulomb centers, it is convenient
to use the elliptic coordinate system ($\xi$, $\eta$):
\begin{equation}
\xi\equiv\frac{r_1+r_2}{R}, \ \  \eta\equiv\frac{r_1-r_2}{R},
\end{equation}
where $R$ is the distance between the two Coulomb impurities, $\xi$ takes values $1\leq\xi<\infty$, and $\eta$ takes values in the interval
$-1\leq\eta\leq 1$. The impurity positions correspond to the points
$\xi=1,\eta=\pm 1$. We note that the elliptic coordinate system is the standard approach to solve the two Coulomb centers problem.\cite{Popov}
The interaction potential in this coordinate system has the form
\begin{equation}
V\left(\mathbf{r}\right)=
\begin{cases}
-\frac{2\zeta v_F \xi}{R(\xi^2-\eta^2)},&\text{\,\,\,\,\, \( \xi^2-\eta^2>4R_0/R\), }\\
-\frac{\zeta v_F}{2R_0}, &\text{\,\,\,\,\, \(\xi^2-\eta^2<4R_0/R\).}
\end{cases}
\end{equation}
We assume that $R \gg R_0$ and if the distance between the electron and an impurity is
less than $R_0$ we neglect the potential due the other impurity. To find the asymptotic of $\phi$
in the vicinity of impurities, i.e., for small $\xi^2-\eta^2$, we seek for a function $\phi$ in the form
$\phi(\xi,\eta)=\phi(\mu)$, where $\mu=\xi^2-\eta^2=4r_1r_2/R^2$.
Near the impurities, i.e., for $\xi\rightarrow 1$ and $\eta\rightarrow \pm 1$ and, consequently,
$\mu \to 0$, we obtain the following equation:
\begin{equation}
\frac{d^2\phi}{d\mu^2}+\frac{1}{\mu}\frac{d\phi}{d\mu}+\frac{R^2\zeta^2}{64R^2_0}\phi=0,
\label{eq:mu-to-zero}
\end{equation}
whose solution regular at $\mu \to 0$ is given by
\begin{equation}
\phi_{\rm imp}(\mu)=CJ_0\left(\frac{\zeta R\mu}{8R_0}\right),
\label{asymptotic-center}
\end{equation}
where $J_0(x)$ is the Bessel function. We note that for charges of impurities such as
$Z\alpha/\kappa< 1/2$ ($\zeta<1$) there is no ``collapse" in the  Coulomb
field of one impurity,\cite{Shytov,Pereira,excitonic-instability,Khalilov,Gupta} therefore,
it is not necessary to cut off the potential at small $r$ and the impurities may be
considered as pointlike. Since $R_0$ little affects the results for
subcritical charges, for simplicity, in what follows, we will consider the
nonregularized Coulomb potential ($R_0=0$). Then instead of Eq.~(\ref{eq:mu-to-zero}) we get
\begin{equation}
\frac{d^2\phi}{d\mu^2}+\frac{2}{\mu}\frac{d\phi}{d\mu}+\frac{\zeta^2}{4\mu^2}\phi=0,
\end{equation}
whose regular solution at $\mu \to 0$ is
\begin{equation}
\phi_{\rm imp}(\mu)=C_2\mu^{-\sigma/2},\quad \sigma=1-\sqrt{1-\zeta^2}.
\label{phi-mu-to-zero}
\end{equation}
This asymptotic describes the behavior of the wave function  at the impurities
positions. Since at large distances $m r>>1$ the variable $\mu$ equals $\mu\simeq
4r^2/R^2$, the solution (\ref{phi-r-to-infinity}) can be rewritten as follows:
\begin{equation}
\phi(\mu)=C_1\mu^{-1/4}K_{i\gamma}(2\sqrt{m\zeta R}\mu^{1/4}).
\label{phi-mu-to-infinity}
\end{equation}
Matching solutions (\ref{phi-mu-to-zero}) and (\ref{phi-mu-to-infinity}) at the point
$\mu=1$ we can find an approximate estimate of the critical distance $R_{\rm cr}(\zeta)$
as a function of $\zeta$. We obtain the following transcendental equation:
\begin{equation}
2\sqrt{1-\zeta^2}-1=2\sqrt{m\zeta R}\frac{K_{i\gamma}^\prime(2\sqrt{m\zeta R})}{K_{i\gamma}
(2\sqrt{m\zeta R})}.
\label{trancend-eq}
\end{equation}
For $m R<<1$, i.e., when the distance between the impurities is much less than the Compton
wavelength of quasiparticles, Eq.~(\ref{trancend-eq}) can be simplified using the asymptotic of $K_{i\gamma}(z)$ for $z \to 0$. Then we obtain the following analytical solution:
\begin{equation}
mR_{\rm cr}=\frac{1}{\zeta}\exp\left[-\frac{2}{\gamma}\left(\cot^{-1}\frac{1-2\sqrt{1-\zeta^2}}{\gamma}
-{\rm arg}\Gamma(1+i\gamma)\right)\right],
\label{appr-num-solution}
\end{equation}
where $\Gamma(z)$ is the Euler gamma function. It is amazing that Eq.~(\ref{appr-num-solution}) coincides with that obtained in QED
for scalar particles.\cite{Popov-Jetp-Lett} Eq.~(\ref{appr-num-solution}) for
$\sqrt{4\zeta^2-1}<<1$  can be written even in more simple form,
\begin{equation}
mR_{\rm cr}=\frac{1}{\zeta}\exp\left(-\frac{2\pi}{\sqrt{4\zeta^2-1}}\right).
\label{critical-distance-estimate}
\end{equation}
We find that the deviation of $R_{\rm cr}$ given by Eq.~(\ref{critical-distance-estimate})
from that determined by Eq.~(\ref{trancend-eq}) is rather small up to $\zeta=0.8$. A numerical
calculation of $R_{\rm cr}$ given by these equations is presented in Fig. \ref{appr-crit-line} in
comparison with $R_{\rm cr}$ determined in more refined calculations using a variational method in Sec.~\ref{variation-appr}.
Clearly, the approximation we used in this section is rather crude  because
it matches only the asymptotics and, in particular, it does not take into account at all the
nonsphericity of the potential of two impurities described by $P_2(\cos\theta)$ and higher
harmonics in potential (\ref{potential-asymptotic}). In Sec. \ref{variation-appr} we present a
more elaborated method for calculating the critical distance $R_{\rm cr}$ following an approach
already successfully used in QED. Before we proceed with this method we will determine in
the next section the energy and width of a quasistationary state present in the system in
the supercritical regime $R<R_{\rm cr}$.

\section{Quasistationary state}
\label{QS-state}

In this section we study a quasistationary state in graphene with two charged impurities and
determine its energy and width. Since an analytic solution for quasistationary states cannot
be found and even the variational method considered in Sec.~\ref{variation-appr} cannot be utilized,
we will use the Wentzel--Kramers--Brillouin (WKB) method. A direct application of the WKB
method to many-body systems which do not admit separation of variables is a complicated
problem because it requires solving the corresponding partial differential equation. Therefore,
we will follow in our analysis Refs.~[\onlinecite{Voskresenskii,Popov-review}], where the WKB method
in the monopole approximation was used in the study of the two-center problem in QED. Although we
need to consider only the gapless case, for the sake of generality, we will deduce the main equations
in the case $\Delta \ne 0$.

For distances $r > R$ (or more exactly $r \gg R$), the potential of the two-center problem is close
to spherically symmetrical one. Therefore, we can consider one charged impurity with the
charge $2Ze$ and restrict our consideration only to the region $r \ge \varkappa R$, where $\varkappa
\sim 1$ is a dimensionless constant. This approximation is known as the monopole approximation.
\cite{RafSoff} We will see that all our results for the energy and width of quasistationary states
practically will not depend on the exact value of $\varkappa$.

For a spherically symmetric potential, we seek eigenfunctions of Eq.~(\ref{Dirac-equation}) in the
following form:
\begin{equation}
\label{sp}
\Psi(\mathbf{r})=\frac{1}{2\sqrt{\pi  r}}\left(
\begin{array}{c}
(1+i) F (r) e^{i (j-1/2) \varphi } \\
(1-i) G (r) e^{i (j+1/2) \varphi } \\
\end{array}
\right),
\end{equation}
where $j=\pm 1/2, \pm 3/2, ...$ is the total angular momentum. Then Eq.~(\ref{Dirac-equation}) reduces to
\begin{equation}
\label{sys}
\left(
\begin{array}{c}
 F'(r) \\
 G'(r) \\
\end{array}
\right)=\frac{1}{\hbar  \upsilon _F}
 \left(
\begin{array}{cc}
 \frac{\tilde{j}}{r} & \Delta +E-V(r)  \\
 \Delta -E+V(r)  & -\frac{\tilde{j}}{r} \\
\end{array}
\right) \left(
\begin{array}{c}
 F(r) \\
 G(r) \\
\end{array}
\right),
\end{equation}
where $\tilde{j}=\hbar v_F j$ and we restore in this section the Planck constant $\hbar$. Making the substitutions
\begin{equation}
\chi_1(r)=(E+\Delta-V(r))^{-1/2} F(r), \ \ \chi_2(r)=(E-\Delta-V(r))^{-1/2} G(r)
\end{equation}
and expressing $\chi_2$ in terms of $\chi_1$, we obtain the following second-order differential
equation for $\chi\equiv\chi_1$:
\begin{equation}
\label{Schr}
\chi''+\frac{E^2-\Delta^2-U}{\hbar^2v^2_F}\chi=0,
\end{equation}
where
\begin{equation}
\label{U}
U(r,E)=2EV-V^2+\hbar^2v_F^2\left(\frac{j(j-1)}{r^2}+ \frac{V''}{2W}+\frac{3}{4}\frac{V'^2}{W^2}
+\frac{j}{r} \frac{V'}{W}\right),
\end{equation}
and $W=E+\Delta-V(r)$.

According to the WKB method, Eq.~(\ref{Schr}) implies the following quasiclassical momentum
for radial motion:
\begin{equation}
p(r,E)=\frac{1}{v_F}\sqrt{E^2-\Delta^2-U(r,E)}.
\end{equation}
Since we are interested in the case $E\lesssim-\Delta$, we introduce a dimensionless parameter
$y=-{(E+\Delta)r}/{(\hbar v_F \zeta)}$,
which is small in the mentioned region. Further, $W=\frac{\hbar \upsilon_F \zeta}{r}(1-y)$.
In order to improve the accuracy of quasiclassical analysis in the region of small $r$, we
introduce the Langer correction making the replacement $j(j-1) \to (j-1/2)^2$. Then, for
the quasiclassical momentum, we have
\begin{equation}
p(r)=\frac{\hbar}{r}\left[\zeta^2-(j-1/2)^2+\frac{2E\zeta r}{\hbar v_F}+
\frac{(E^2-\Delta^2)r^2}{\hbar^2 v_F^2}+\frac{1-j}{1-y}
-\frac{3}{4 (1-y)^2}\right]^{1/2}.
\label{quasiclassical-momentum}
\end{equation}
Expanding the quasiclassical momentum (\ref{quasiclassical-momentum}) in series in $y$ and retaining
terms up to $y^2$, we find
\begin{equation}
\label{asym}
p(r)\simeq\frac{\hbar}{r}\left\{a- 2 b r+c r^2 + O\left[(E+\Delta)^3\right]\right\}^{1/2},
\end{equation}
where the  coefficients $a,b,c$ are
\begin{equation}
\label{a}
a=\zeta^2-j^2, \quad b=\frac{\zeta}{\hbar v_F}\left[\Delta-\left(1+\frac{1+2j}{4\zeta^2}\right)
(E+\Delta)\right],\quad c=\frac{E^2-\Delta^2}{\hbar^2 v_F^2}-\frac{(j+5/4)(E+\Delta)^2}{\hbar^2 v_F^2 \zeta^2}.
\end{equation}
For energies near the boundary $E\lesssim-\Delta$ and $\zeta>|j|$ these coefficients are
positive and a classically forbidden region is defined by $r_{-}<r<r_{+}$  where $p^2(r)$ is negative.
The turning points $r_{\pm}=(b\pm\sqrt{b^2-a c})/c$ are determined by the equation $p(r_{\pm},E)=0$
and depend on the energy $E$. The quasiparticles with wavelength less than
$r_{-}$ can be trapped in the region $r<r_{-}$ and their lifetime is defined by tunneling through
the barrier.
To find the energy of quasibound states we use the Bohr--Sommerfeld quantization condition
\begin{equation}
\int\limits_{\varkappa R/2}^{r_{-}}p(r,E)dr=\pi\hbar n,
\label{Bohr-Somm}
\end{equation}
where $n=1,2...$. The lower cutoff $\varkappa R/2$ is related to the size of quasimolecule and
$\varkappa$ is a numerical factor of order of one which defines the accuracy of the considered
monopole approximation.
Since for $E=-\Delta$ and $n=1$ Eq.~(\ref{Bohr-Somm}) becomes an equation for $R_{\rm cr}$, the equation
for $E(R)$ of the state with $n=1$ takes the form
\begin{equation}
\int\limits_{\varkappa R/2}^{r_{-}}p(r,E)dr=\int\limits_{\varkappa R_{\rm cr}/2}^{r_{-}^0}
p(r,-\Delta)dr,
\label{Bohr-Somm1}
\end{equation}
where $r_{-}^0=r_{-}(E=-\Delta)$. The integration in Eq.~(\ref{Bohr-Somm1}) can be performed
in explicit form (see the Appendix). The critical distance $R_{\rm cr}$ is determined from Eq.~(\ref{Bohr-Somm})
for $E=-\Delta$ and $n=1$, and is given by
\begin{equation}
\frac{R_{\rm cr}\Delta}{\hbar v_F}=\frac{4(\zeta^2-j^2)}{\zeta\varkappa}\exp
\left(-\frac{\pi}{\sqrt{\zeta^2-j^2}}-2\right),\quad \zeta>|j|.
\label{Rcr}
\end{equation}
For energies close to the boundary of the lower continuum, $E\to-\Delta$,
we find from Eq.~(\ref{eq:Rcr}),
\begin{equation}
{E(R,\zeta)}=-{\Delta}\cdot F(\zeta,R), \quad\quad F(\zeta,R)=\left(\frac{R_{\rm cr}}{R}+\frac{1+2j}{4\zeta^2}-\frac{\zeta^2-j^2}{3\zeta^2}\right)/
\left(1+\frac{1+2j}{4\zeta^2}-\frac{\zeta^2-j^2}{3\zeta^2}\right).
\label{energy-resonance}
\end{equation}
Clearly, $F(\zeta,R=R_{\rm cr})=1$, where $R_{\rm cr}$ is given by Eq.~(\ref{Rcr}).
In particular, one can see that for the lowest-energy state with $j=1/2$, $R_{\rm cr}$ tends to infinity
as the gap $\Delta \to 0$ if $\zeta > 1/2$. Since graphene is gapless in the absence of external
fields, this result suggests that two charged impurities are always in the supercritical regime
as soon as their total charge exceeds the critical one $\zeta_{\rm c} = 1/2$.

The width of quasistationary states apart from a preexponential factor is determined
by tunneling through the classically forbidden region,
\begin{equation}
\Gamma\propto\exp\left(-2\int\limits_{r_{-}}^{r_{+}}\frac{\sqrt{-a+2b r- c r^2}}{r}dr\right)
=\exp\left(-2\pi\left(\frac{b}{\sqrt{c}}-\sqrt{a}\right)\right).
\end{equation}
For energies close to the boundary of the lower continuum this gives the width
\begin{eqnarray}
\Gamma\propto \exp\hspace{-1mm}\left[-2\pi\hspace{-1mm}\left(\zeta\sqrt{\frac{E^2}{E^2-\Delta^2}}
-\sqrt{\zeta^2-j^2}\right)\hspace{-1mm}\right]\simeq\exp\hspace{-1mm}\left[-2\pi\hspace{-1mm}
\left(\tilde\beta\sqrt{\frac{R}{R_{\rm cr}-R}}-\sqrt{\zeta^2-j^2}\right)\hspace{-1mm}\right],
\tilde\beta=\sqrt{\frac{8\zeta^2+4j^2+6j+3}{24}},
\label{width-Gamma}
\end{eqnarray}
which tends to zero when $E\to-\Delta$ or $R\to R_{\rm cr}$.

In the case of gapless quasiparticles, the formulas simplify and we get for the energies
of quasistationary states with $\zeta\to |j|$ the expression (\ref{Rcr})  where one should make
replacements $\Delta\to|E|(1+(1+2j)/4\zeta^2)$ and $R_{\rm cr}\to R$. The width of this state is given
by Eq.~(\ref{width-Gamma}) for $\Delta=0$. In contrast to the case of gapped quasiparticles,
the width of quasistationary states in gapless graphene has no energy dependence.

\section{Variational method }
\label{variation-appr}

The nonrelativistic Schr\"odinger equation for the electron in the potential of two Coulomb
centers permits separation of variables in elliptic coordinates. Therefore, it is an analytically
solvable problem and is extensively used in the theory of chemical binding. Unfortunately, as we mentioned
above, for the Dirac equation, variables are not separable in any known orthogonal coordinate
system and it is not possible to obtain its solution in an analytic form. Therefore, in order
to study the supercritical instability of two Coulomb centers in graphene we will employ as in QED\cite{Popov} the variational method. As noted in Ref.~[\onlinecite{variational}], to obtain
a satisfactory accuracy it is necessary that trial functions correctly reproduce the asymptotics
of the exact solution at infinity and near the charged impurities. These asymptotics in the case
under consideration are given by Eqs.~(\ref{asymptotic}) and (\ref{phi-mu-to-zero}), respectively.

To set up the variational problem, we note that the differential equation~(\ref{upper-component})
can be obtained as an extremum of the following functional:
\begin{equation}
S[\phi]=\int\left((E-V+\Delta)^{-1}\left|\frac{\partial\phi}{\partial x}+i\frac{\partial\phi}
{\partial y}\right|^2-v^{-2}_F(E-V-\Delta)|\phi|^2\right) dxdy,
\label{functional}
\end{equation}
under the condition that the norm $N=\int\Psi^*\Psi dxdy$ is conserved (the norm is important for obtaining the correct boundary conditions).
Introducing new field $\psi=W^{-1/2}\phi, W=E-V+\Delta$ the functional $S[\phi]$ can be represented
in the form specific for nonrelativistic quantum mechanics,
\begin{equation}
S[\psi]=\int\left[|\bm\nabla\psi|^2+i\left(\frac{\bm\nabla V}{2W}\times\bm\nabla\psi^*\right)
\psi-i\psi^*\left(\frac{\bm\nabla V}{2W}\times\bm\nabla\psi\right)+2(U-\epsilon)|\psi|^2\right]dxdy,
\label{functionalS-modified}
\end{equation}
where $\mathbf{a}\times\mathbf{b}=\epsilon_{ij}a_ib_j$, $\epsilon=(E^2-\Delta^2)/2v_F^2$ is
the effective energy, and the effective potential $U$ is given by
\begin{equation}
U=\frac{EV}{v_F}-\frac{V^2}{2v_F^2}+\frac{\triangle V}{4W}+\frac{3}{8}\frac{(\bm\nabla V)^2}{W^2}.
\end{equation}
The second and third terms in functional (\ref{functionalS-modified}) describe the pseudospin-orbit
coupling with the field $\mathbf{F}=-{\bm\nabla V}/2W$, they  do not contribute for the ground
state wave function which is real.
Functional (\ref{functionalS-modified})
is bounded from below, so one is in position to apply to it the variational principle. In what follows
we are interested in the case where the bound state with the lowest energy crosses the boundary of
the lower continuum, so we put $E=-\Delta$ ($\epsilon=0$). Then $W=-V$ and the functional $S[\psi]$
is simplified.

In QED, the Ritz and Kantorovich methods were employed in order to solve the variational problem
and find a critical distance $R_{\rm cr}$ (see a discussion in Sec. III in Ref.~[\onlinecite{Popov-review}]).
In the Ritz method, the sought function $\psi$ is expanded over a fixed set of basis
functions $\psi(x,y)=\sum_nc_n\psi_n(x,y)$, where $c_n$ are variable constants. In the Kantorovich
method, $\psi=\sum_nc_n(x)\psi_n(y)$, where $\psi_n(y)$ are fixed functions, while $c_n(x)$ are
variable functions. Obviously, the variational problem reduces to a system of linear algebraic
equations for $c_n$ in the Ritz method and to a system of linear ordinary differential
equations for $c_n(x)$ in the Kantorovich method.

According to Eq.~(\ref{phi-mu-to-zero}), near impurities $\phi$ depends only on $\mu=\xi^2-\eta^2
={4r_1 r_2}/{R^2}$. At the large distances, $r\to\infty,$ the variable $\mu \to \infty$ and the asymptotic of $\phi$ is given by Eq.~(\ref{phi-mu-to-infinity}). Therefore, both asymptotics of $\psi$ depend only on $\mu$. In order that a variational ansatz for $\psi$ gives appropriate results, it is essential to take into account correctly the behavior of the exact solution near the Coulomb centers and at infinity. We choose the variables $\mu,\nu$ so
that the function $\psi(x,y)$ has a singularity only in $\mu$. Then
using the following ansatz in the Kantorovich method,
\begin{equation}
\psi=\sum\limits_{k=1}^{N}\psi_k(\mu)\nu^{k-1},
\label{Kantorovich-method}
\end{equation}
where $\psi_k(\mu)$ are variable functions of $\mu$ and $\nu(\xi,\eta)$ is a fixed function of $\xi$
and $\eta$, we can maximally correctly take into account the behavior of the exact solution near
the Coulomb centers. Since {\it a priori} we do not know what set of functions $\psi_n(x,y)$ is
the best in the Ritz method, in the present paper like in QED studies\cite{Popov} we will use the Kantorovich method.

Two variants for $\nu$ were considered in QED:\cite{Popov}
i) $\nu=\eta^2/(\xi^2-\eta^2)=(r_1-r_2)^2/4r_1r_2$ and ii) $\nu=\eta^2=(r_1-r_2)^2/R$.
The results obtained were close. In this paper, we will consider the case i).
Since the charges of impurities are identical, $Z_1=Z_2=Z$, the wave function of the ground
state is symmetric under the inversion $x\rightarrow -x, y\rightarrow -y$, therefore,
the change of the variables $x,y$ to $\mu,\nu$  is performed  by means of the formulas,
\begin{equation}
x=\frac{R}{2}\mu\sqrt{\nu(\nu+1)},\quad y=\frac{R}{2}[(\mu(\nu+1)-1)(1-\mu\nu)]^{1/2}.
\end{equation}

Inserting ansatz (\ref{Kantorovich-method}) in Eq.~(\ref{functionalS-modified}) and
integrating over $\nu$, we obtain
\begin{equation}
S_N(\psi)=4\sum\limits_{k,l=1}^{N}\int\limits_{0}^{\infty}d\mu \left(P_{kl}\psi_k'{\psi^*_l}'
+Q_{kl}\psi_k\psi^*_l+R_{kl}\psi_k'\psi^*_l+R_{kl}^{\dagger}{\psi_k}{\psi^*_l}'\right),
\label{functional2}
\end{equation}
where $P,Q$, and $R$ are $N\times N$ matrices which depend on $\mu$ and are given by
Eqs.~(\ref{P})-(\ref{R}) in Appendix.
A formula similar to functional~(\ref{functional2}) may be also obtained for the norm.

Minima of functional~(\ref{functional2}) are given by solutions of the following set of
Euler-Lagrange equations:
\begin{equation}
\frac{d}{d\mu}\left(P_{kl}\frac{d\psi_k}{d\mu}+R_{kl}^{\dagger}\psi_k\right)-Q_{kl}\psi_k-R_{kl}
\frac{d\psi_k}{d\mu}=0.
\label{set-of-equations}
\end{equation}
The boundary conditions for functions $\psi_k$ follow from the requirement that the norm of
the function $\psi$ be finite. The differential equation~(\ref{set-of-equations}) and these
boundary conditions define our boundary value problem. In the simplest case $N=1$, we
have
\begin{equation}
\label{e4}
\frac{d}{d\mu}\left(P\frac{d\psi}{d\mu}\right)-Q\psi=0,
\end{equation}
where $P(\mu)=\pi\mu$ and $Q(\mu)$ is expressed through the complete elliptic integrals
of the first and second kind (see, Eq.~(\ref{Q-function}) in Appendix) and has a logarithmic
singularity at $\mu=1$. Asymptotics of the function $Q(\mu)$ at small and large values of
$\mu$ are given by the expressions
\begin{eqnarray}
Q(\mu)\simeq\left\{\begin{array}{c}\frac{\pi(1-\zeta^2)}{4\mu},\quad \mu\to 0,\\
\pi\left(\frac{\zeta v_F m R}{4\sqrt{\mu}}+\frac{1-4\zeta^2}{16\mu}\right),\quad\mu\to\infty.
\end{array}\right.
\end{eqnarray}
Taking into account these asymptotics, Eq.~(\ref{e4}) can be solved analytically in the regions
$\mu<1$ and $\mu>1$. The corresponding solutions regular at $\mu=0$ and decreasing at
$\mu \to \infty$ are
\begin{equation}
\psi(\mu)=C_1\mu^{\sqrt{1-\zeta^2}/2},\quad\quad \mu<1,
\label{sol-small-mu}
\end{equation}
\begin{equation}
\psi(\mu)=C_2K_{i\gamma}\left(2\sqrt{\zeta mR}{\mu}^{1/4}\right),\quad\gamma=\sqrt{4\zeta^2-1},
\quad\quad \mu>1.
\label{sol-large-mu}
\end{equation}
These asymptotic solutions are in agreement with Eqs.~(\ref{phi-mu-to-zero}) and (\ref{phi-mu-to-infinity}).
The substitution $\psi=\mu^{-1/2}\chi$ recasts Eq.~(\ref{e4}) in the form of Schr{\"o}dinger-like
equation for zero energy
\begin{equation}
-\chi''(\mu)+V_{\rm eff}(\mu)\chi(\mu)=0, \quad V_{\rm eff}(\mu)=-\frac{1}{4\mu^2}+\frac{Q(\mu)}{\pi\mu}.
\end{equation}
The effective potential $V_{\rm eff}(\mu)$ has a wide positive barrier due to the term proportional to
$\zeta mR$ that explains the exponential decreasing of the wave function (\ref{sol-large-mu})
at large $\mu$.

The differential equation (\ref{e4}) determines the wave function of the critical bound state that just dives into the lower continuum. Since
the wave function of a bound state tends to zero at infinity, this translates in our case to the condition $\psi(\mu) \to 0$ as $\mu \to
\infty$. The asymptotic of the wave function near the impurities (where $\mu \to 0$) is given by Eq.~(\ref{sol-small-mu}). This equation
completes the set-up of our boundary value problem which allows us to determine the critical distance $R_{\rm cr}$ between the impurities as a
function of $\zeta$. Since the function $Q(\mu)$ is given in terms
of the complete elliptic integrals of the first and second kind, the differential equation (\ref{e4})
cannot be solved analytically. We solve this equation numerically by using the shooting method and proceed
as follows. We fix the wave function and its first derivative at certain small $\mu$  using Eq.~(\ref{sol-small-mu}). Note that since the
differential equation (\ref{e4}) is linear, the value of the normalization constant $C_1$ is irrelevant. Therefore, for simplicity, we choose
$C_1=1$. Furthermore, we fix $\zeta$ and solve Eq.~(\ref{e4}) numerically by using {\it Mathematica} for different $mR$ [note that
since the function $Q(\mu)$ depends only on the product $mR$, parameters $m$ and $R$ cannot be separately varied]. The critical distance
$R_{\rm cr}$ (for a given $m$) is then determined as $R$ such that the wave function $\psi(\mu)$ tends to zero at infinity. Repeating this procedure
for different $\zeta$, we find
how the critical distance between the impurities depends on $\zeta$. The corresponding
dependence $mR_{\rm cr}$ on $\zeta$ is plotted in Fig.~\ref{appr-crit-line} (solid red line).

The accuracy of computation can be improved taking $N>1$ in the sum (\ref{Kantorovich-method}).
In this case one should solve a set of second-order differential equations. Since the shooting method
is not well suited for this purpose, it is better then to follow analogous calculations in QED in Ref.~[\onlinecite{Marinov}] and reduce the set of Eqs.~(\ref{set-of-equations}) to the matrix Riccati equation, which can be solved by the Runge-Kutta method.

\section{Conclusion}
\label{section-conclusion}

Motivated by a recent observation of atomic collapse in clusters of four and five charged Ca dimers in graphene, we studied  the supercritical instability in the simplest cluster of charged impurities in graphene formed by two similar impurities whose charges are subcritical like in the experiment. It is possible that future experiments in suspended graphene where the screening due to the substrate is absent may observe the supercritical instability directly in the system of two Coulomb impurities. In our study we assumed that the total charge of two impurities $2Ze$ exceeds the critical charge (determined by the condition $\zeta_{\rm c}=2Z_{\rm c}\alpha/\kappa=1/2$) if these impurities are placed together. Therefore, at fixed $\zeta$ the supercritical regime sets in for a certain  critical distance $R_{\rm cr}$ between the impurities.

Since the variables in the Dirac problem with two Colulomb centers are not separable in any known
orthogonal coordinate system, this problem does not admit an analytic solution. Therefore, in order
to find the dependence of $R_{\rm cr}$ on $\zeta$ we used the variational Kantorovich method.
For gapless quasiparticles, the supercritical instability is signaled by the appearance of resonances.
Since it is difficult to study resonances formed by gapless quasiparticles in a variational method,
we introduced a small gap $\Delta=mv^2_F$ for quasiparticles and looked for the electron bound state
with lowest energy that dives into the lower continuum. The distance between the impurities when this
happens defines $R_{\rm cr}$.

The dependence of $mR_{\rm cr}$ on $\zeta$ is plotted in Fig.~\ref{appr-crit-line} together
with approximate analytical solutions obtained in the monopole approximation.
Naturally, the critical distance, separating the supercritical, $R < R_{\rm cr}$, and subcritical,
$R > R_{\rm cr}$ regimes, tends to zero  as $\zeta \to 1/2$ and $R_{\rm cr} \to \infty$ as
$\zeta \to 1$ (in the last case the charge of each impurity tends to the critical one).  It means
that the system is always in the subcritical  regime if the total charge is less than the critical
one $\zeta_{\rm c}=1/2$, and in the supercritical regime if the charge of each  impurity is larger than
the critical charge. Our results show that at fixed $\zeta$ the critical distance tends to
infinity for $m \to 0$. This means that in the considered model, as soon as the total charge
of two impurities exceeds the critical one $\zeta_{\rm c}=1/2$, the system for gapless quasiparticles
is in the supercritical regime for any distance between the impurities.

In the real specimen, there is always a remnant density of charge carriers that screens
the Coulomb potential. The Thomas--Fermi screening wave vector in graphene equals
$q_{\rm TF}=4\pi^{1/2}\alpha\sqrt{n}/\kappa$ and for distances that exceed $l_{\rm TF}=1/q_{\rm TF}$
the Coulomb interaction is screened. In this case the critical distance for gapless quasiparticles
and charges $1/2<\zeta<1$ is defined by the Thomas--Fermi screening length $l_{\rm TF}$.
According to Ref.~[\onlinecite{Mayorov}], the lowest charge density inhomogeneity attainable
at present experimentally is $n\approx 10^{8}$  cm$^{-2}$. Therefore, we find that the model
with the Coulomb interaction can be used if the distance between impurities is less than
$100\div 400$ nm. For less clean samples with the density $n\approx 10^{10}$  cm$^{-2}$ the length
$l_{\rm TF}$ is one order smaller. Since the distance between calcium dimers in the experiment~[\onlinecite{Wang}] is $d\sim 2$ nm, we conclude that for the individual impurities the
Coulomb interaction can be used.

In the present paper we studied the instability in graphene with two charged impurities while
the experiment\cite{Wang} deals with clusters of four and five impurities. Clearly, the study
of such clusters can be done only numerically, except the simplest monopole approximation, and
this is a challenge for future investigations.

\begin{acknowledgments}
We thank V.A.~Miransky and I.A.~Shovkovy for useful remarks.
This work is supported partially by the European FP7 program, Grant No. SIMTECH
246937, the joint Ukrainian-Russian SFFR-RFBR Grant No.~F53.2/028, the grant STCU \#5716-2
"Development of Graphene Technologies and Investigation of Graphene-based Nanostructures for
Nanoelectronics and Optoelectronics", and by the Program of Fundamental Research of the Physics
and Astronomy Division of the NAS of Ukraine. V.P.G. acknowledges a collaborative
grant from the Swedish Institute.
\end{acknowledgments}

\appendix
\section{}
\label{A}

In this Appendix we perform the integration in Eq.~(\ref{Bohr-Somm1}) and derive the expressions for
the matrices $P,Q,R$ in the functional (\ref{functional2}). The integration in Eq.~(\ref{Bohr-Somm1})
can be performed exactly but the corresponding expression is more transparent for small $r$,
\begin{equation}
\int\limits_{r}^{r_{-}}\frac{\sqrt{a-2b r+c r^2}}{r}dr=\sqrt{a}\left[\ln\frac{2a}{b r}
-f(x)-2\right]+O(r),\quad r<<r_{-},
\quad x=\frac{{a c}}{b^2},
\end{equation}
where the function
\begin{eqnarray}
f(x)=\left\{\begin{array}{c}\frac{1}{2}\ln(1-x)+\frac{1}{2\sqrt{x}}\ln\frac{1+\sqrt{x}}{1-\sqrt{x}}-1,
\quad x>0,\\ \frac{1}{2}\ln(1-x)+\frac{\arctan\sqrt{|x|}}{\sqrt{|x|}}-1,\quad x<0.
\end{array}\right.
\end{eqnarray}
Then Eq.~(\ref{Bohr-Somm1}) can be written in the form
\begin{equation}
\frac{R_{\rm cr}}{R}=\left[1-\left(1+\frac{1+2j}{4\zeta^2}\right)\frac{E+\Delta}{\Delta}\right]e^{f(x)},
\label{eq:Rcr}
\end{equation}
where
\begin{equation}
 x=\frac{\zeta^2-j^2}{\zeta^2}\left[E^2-\Delta^2-\frac{4j+5}{4\zeta^2}(E+\Delta)^2\right]
\left[\Delta-\left(1+\frac{1+2j}{4\zeta^2}\right)(E+\Delta)\right]^{-2}.
\end{equation}
For energies close to the boundary of the lower continuum, $E\to-\Delta$, the variable $x\to 0$
and we come to Eq.~(\ref{energy-resonance}) in the main text.

Now we derive the expressions for the matrices $P,Q,R$ in the functional (\ref{functional2}).
The functional (\ref{functionalS-modified}) for $E=-\Delta$ takes the form
\begin{eqnarray}
S[\psi]&=&4\sum\limits_{k,l=1}^{N}\int\limits_0^\infty d\mu d\nu|J|\left[(\bm\nabla\mu)^2\psi_{k}'{\psi^*_l}'\nu^{k+l-2}
+2\bm\nabla\mu\bm\nabla\nu\Re e(\psi^*_l\psi_k')(l-1)\nu^{k+l-3}-2\left(\frac{\bm\nabla V}{2V}\times\bm\nabla\mu\right)\Im m(\psi^*_l\psi_k')\nu^{k+l-2}\right.\nonumber\\
&+&\left.\psi^*_l\psi_k[(\bm\nabla\nu)^2(l-1)(k-1)
\nu^{k+l-4}-i(l-k)\left(\frac{\bm\nabla V}{2V}\times\bm\nabla\nu\right)\nu^{k+l-3}+2U\nu^{k+l-2}]\right]
f(\mu,\nu),
\label{functional-1}
\end{eqnarray}
where the functions $\bm\nabla\mu,\bm\nabla\nu,V,U$ should be expressed through the variables
$\mu,\nu$. Note that for the ground state wave function which is real the third term in Eq.~(\ref{functional-1}) does not contribute.
Since $\mu\nu=\eta^2<1,\quad \mu(\nu+1)=\xi^2>1$,
the integration in the $(\mu,\nu)$ plane is performed over the curvilinear triangle,
\begin{equation}
\left(\frac{1}{\mu}-1\right)\theta\left(1-\mu\right)<\nu<\frac{1}{\mu},
\end{equation}
that is provided by the function $f(\mu,\nu)$,
\begin{equation}
f(\mu,\nu)=\theta(1-\mu\nu)[\theta(1-\mu)\theta(\mu(\nu+1)-1)+\theta(\mu-1)].
\end{equation}
We find
\begin{eqnarray}
&&|J|=\frac{\mu R^2}{16}\frac{1}{\sqrt{\nu(\nu+1)(\mu+\mu\nu-1)(1-\mu\nu)}}\,,
\quad V(\mu,\nu)=-\frac{2v_F\zeta}{R}\sqrt{\frac{\nu+1}{\mu}},\\
&&(\bm\nabla\mu)^2=\frac{16}{R^2}(\mu+2\mu\nu-1),\quad\quad (\bm\nabla\nu)^2=\frac{16\nu(\nu+1)}{\mu^2 R^2}, \quad\quad \bm\nabla\mu\bm\nabla\nu=-\frac{16\nu(\nu+1)}{R^2},\\
&&\frac{\bm\nabla V}{V}\times\bm\nabla\mu=(\bm\nabla\nu\times\bm\nabla\mu)\frac{\partial\ln V}{\partial\nu}
=\frac{8}{R^2\mu}\frac{\sqrt{\nu(\mu+\mu\nu-1)(1-\mu\nu)}}{\sqrt{\nu+1}},\\
&&\frac{\bm\nabla V}{V}\times\bm\nabla\nu=(\bm\nabla\mu\times\bm\nabla\nu)\frac{\partial\ln V}{\partial\mu}
=\frac{8}{R^2\mu^2}\sqrt{\nu(\nu+1)(\mu+\mu\nu-1)(1-\mu\nu)},\\
&&2U=\frac{2}{R^2}\left[2v_F\zeta mR\sqrt{\frac{\nu+1}{\mu}}-2\zeta^2\frac{\nu+1}{\mu}-\frac{4\nu+1}{\mu}+
\frac{3}{2}\frac{4\mu\nu^2+5\mu\nu+\mu-1}{\mu^2(\nu+1)}\right].
\end{eqnarray}
The functional (\ref{functional-1}) takes the form given in Eq.~(\ref{functional2}),
where $P,Q$, and $R$ are $N\times N$ matrices which depend on $\mu$,
\begin{eqnarray}
\label{P}
P_{kl}(\mu)&=&\int\limits_0^\infty (\bm\nabla\mu)^2\nu^{k+l-2}
|J|f(\mu,\nu) d\nu,\\
\label{Q}
Q_{kl}(\mu)&=&\int\limits_0^\infty\left[(\bm\nabla\nu)^2(l-1)(k-1)
\nu^{k+l-4}-i(l-k)\left(\frac{\bm\nabla V}{2V}\times\bm\nabla\nu\right)\nu^{k+l-3}+2U\nu^{k+l-2}\right]
|J|f(\mu,\nu) d\nu , \\
\label{R}
R_{kl}(\mu)&=&\int\limits_0^\infty\left[\bm\nabla\mu\bm\nabla\nu(l-1)\nu^{k+l-3}+
i\left(\frac{\bm\nabla V}{2V}\times \bm\nabla\mu\right)\nu^{k+l-2}\right]|J|f(\mu,\nu) d\nu.
\end{eqnarray}
The second term in Eq.~(\ref{R}) is absent for the ground state wave function.
For $N=1$, we need only the functions $P$ and $Q$,
\begin{equation}
P(\mu)=\mu\int\limits_0^\infty\frac{(\mu+2\mu\nu-1)f(\mu,\nu)}{\sqrt{\nu(\nu+1)(\mu+\mu\nu-1)(1-\mu\nu)}}\,
d\nu,
\end{equation}
\begin{equation}
Q(\mu)=\frac{\mu }{8}\int\limits_0^\infty\frac{f(\mu,\nu)}{\sqrt{\nu(\nu+1)(\mu+\mu\nu-1)(1-\mu\nu)}}\,
\left[2v_F\zeta mR\sqrt{\frac{\nu+1}{\mu}}-2\zeta^2\frac{\nu+1}{\mu}-\frac{4\nu+1}{\mu}+
\frac{3}{2}\frac{4\mu\nu^2+5\mu\nu+\mu-1}{\mu^2(\nu+1)}\right]d\nu.
\end{equation}
These functions can be expressed in terms of elliptic integrals and the needed integrals
are given by Eqs.~(3.131.6), (3.147.6), (3.151.6), (3.167.6), (3.167.22) in Ref.~[\onlinecite{Gradshtein}]. We obtain
\begin{eqnarray}
P(\mu)&=&2\mu\theta(1-\mu)(1-\mu)\left[2\Pi\left(\mu,\mu\right)-K(\mu)
\right]+2\theta(\mu-1)(\mu-1)\left[2\Pi\left(\frac{1}{\mu},\frac{1}{\mu}\right)
-K\left(\frac{1}{\mu}\right)\right],\\
\label{Q-function}
Q(\mu)&=&\frac{v_F\zeta mR}{2}\left[\theta(1-\mu) K(\sqrt{\mu})+\frac{\theta(\mu-1)}{\sqrt{\mu}}K\left(\frac{1}{\sqrt{\mu}}\right)\right]
-\frac{\zeta^2-1}{2\mu}\left[\theta(1-\mu)\left((1-\mu)\Pi(\mu,\mu)+\mu K(\mu)\right)\right.\nonumber\\
&+&\left. \frac{\theta(\mu-1)}{{\mu}}
\left((\mu-1)\Pi\left(\frac{1}{\mu},\frac{1}{\mu}\right)+ K\left(\frac{1}{\mu}\right)\right)
\right]+\frac{3}{8\mu}\left[\theta(1-\mu)\left((1-\mu)\Pi(\mu^2,\mu)-(1+\mu)K(\mu)\right)\right.\nonumber\\
&+&\left.\frac{\theta(\mu-1)}{\mu}\left((\mu-1)\Pi\left(\frac{1}{\mu^2},\frac{1}{\mu}\right)-2\mu
K\left(\frac{1}{\mu}\right)\right)\right],\quad
\end{eqnarray}
where $K(k)$, $E(k)$ and $\Pi(n,k)$ are the complete elliptic integrals of the first, second and third kind, respectively. Using the identities,\cite{Byrd}
\begin{equation}
\Pi(\mu,\mu)=\frac{\pi}{4(1-\mu)}+\frac{1}{2}K(\mu),\quad \Pi(\mu^2,\mu)=\frac{1}{1-\mu^2}E(\mu),
\end{equation}
we find that $P(\mu)=\pi\mu$, while the function $Q(\mu)$ is expressed in terms of the complete
elliptic integrals of the first and second kind.

\end{document}